\documentclass[12pt,nofootinbib]{article}

\usepackage{epsf}
\usepackage{amsmath}
\usepackage{graphicx}
\usepackage{amsfonts}
\usepackage{amssymb}
\usepackage{latexsym}
\usepackage{color}
\usepackage{cite}
\input{colordvi.tex}

\usepackage{feynmp}
\renewcommand{\thefootnote}{\#\arabic{footnote}}

\newcommand{\f}[2]{\frac{#1}{#2}}
\newcommand{\mk}[1]{\left( #1 \right)}
\newcommand{\kk}[1]{\left[ #1 \right]}
\newcommand{\ck}[1]{\left\{ #1 \right\}}

\newcommand{\vac}[1]{\langle 0| #1 |0 \rangle}
\newcommand{\be}{\begin{equation}}
\newcommand{\ee}{\end{equation}}
\newcommand{\bea}{\begin{eqnarray}}
\newcommand{\eea}{\end{eqnarray}}
\newcommand{\fm}[1]{\int \f{d^3 #1}{{(2\pi)}^3}}
\def\del{\partial}
\def\a{\alpha}
\def\d{\delta}
\def\e{\epsilon}
\def\g{\gamma}

\def\x{{\vec x}}

\def\p{{\vec p}}

\def\k{{\vec k}}
\def\l{{\vec \ell}}
\def\h{{\vec h}}
\def\ii{{\vec i}}
\def\j{{\vec j}}
\def\n{{\vec n}}
\def\Hfp{H_{\f{1}{2}+i\a}^{(1)}}
\def\Hfm{H_{\f{1}{2}-i\a}^{(1)}}
\def\Hsp{H_{\f{1}{2}+i\a}^{(2)}}
\def\Hsm{H_{\f{1}{2}-i\a}^{(2)}}

\begin{document}
\setcounter{footnote}{0}

\begin{titlepage}
\begin{flushright}
RESCEU-1/14
\end{flushright}
\begin{center}

\vskip .5in

{\Large \bf
Loop contribution to inflationary magnetic field
}

\vskip .45in

{\large
Hayato Motohashi$^1$
and 
Teruaki Suyama$^2$
}

\vskip .45in%

{\em 
$^{1}$ Kavli Institute for Cosmological Physics, The University of Chicago, 
Chicago, Illinois 60637, U.S.A.
}\\
{\em
$^{2}$ Research Center for the Early Universe (RESCEU),  
Graduate School of Science, The University of Tokyo, Tokyo 113-0033, Japan 
}

\end{center}

\vskip .4in

\begin{abstract}
Within the framework of the standard quantum electrodynamics, 
we compute contribution of vacuum polarization at one-loop order to the power
spectrum of the magnetic field on inflationary (de Sitter) background.
It is found that the one-loop term exhibits the infrared secular growth that 
is proportional to the number of $e$-folds.
The use of the dynamical renormalization group method, which amounts to partial resummation of
higher loop diagrams, shows that the resummed power spectrum is free from the secular
growth and the loop effect only changes the power from $(k/a)^4$ at the tree level to ${(k/a)}^{4-\nu}$, 
where $\nu$ represents the contribution from vacuum polarization.
The parameter $\nu$, being proportional to the square of the gauge coupling constant
as well as the number of fermion species, 
is a simple function of a ratio of fermion mass to the Hubble parameter
and is positive irrespective of the fermion mass.
Thus, the loop effect always enhances the infrared magnetic field strength.
We find that $\nu \simeq 5 \times 10^{-3}$ is the possible maximum contribution
to $\nu$ from a single fermion.
This estimate suggests that either large number of fermion species or large coupling constant is 
a necessary condition for the loop effect to be responsible for the seed of the cosmic magnetic fields.
\end{abstract}
\end{titlepage}

\renewcommand{\thepage}{\arabic{page}}
\setcounter{page}{1}
\renewcommand{\thefootnote}{\#\arabic{footnote}}

\section{Introduction}
Magnetic fields are known to exist ubiquitously at various sites in the Universe
(see for instance \cite{Grasso:2000wj,Widrow:2002ud}).
Measurements of intensity and polarization of synchrotron emission and the Faraday rotation 
reveal that many galaxies and cluster of galaxies possess magnetic fields of the order of 
${\rm \mu G}$ with coherent length comparable to galactic scales or larger.
Observation and non-observation of cosmic gamma-rays place lower bound on 
the intensity of the extragalactic magnetic fields \cite{Neronov:1900zz,Taylor:2011bn}, 
although there is strong degeneracy between magnetic field strength and the coherent length.
Despite these observational supports of the universal existence of the magnetic fields,
no compelling astrophysical scenario can explain their origin.
This fact has led a certain number of cosmologists to seriously look for the solution in 
the primordial inflation.

As pointed out in a paper by Turner and Widrow \cite{Turner:1987bw}, 
inflation provides a natural and simple way of explaining the observed relatively 
large coherent length as a result of enormous stretching of any physical length scales
by inflationary expansion.
However, contrary to the cases of the scalar and the tensor perturbations,
simply putting the Maxwell field on the de Sitter space only results in little amount
of magnetic field on large scales due to the conformal invariance of the Maxwell action.
Three possible ways to break the conformal invariance were investigated in the same paper;
1) gravitational couplings such as $R_{\mu \nu}A^\mu A^\nu$ and $R A_\mu A^\mu$ that also break gauge invariance,
2) gravitational couplings such as 
$R_{\mu \nu \lambda \kappa}F^{\mu \nu}F^{\lambda \kappa},~R_{\mu \nu}F^{\mu \lambda}F^\nu_{~\lambda}$
and $RF_{\mu \nu}F^{\mu \nu}$ that respect gauge invariance,
3) gauge interaction to charged scalar field or coupling to axionic scalar field.
The first possibility is successful in the sense of explaining the observed magnetic fields by 
suitable choice of parameters although dismissal of the gauge invariance makes this option
less attractive.
The couplings in the second possibility produce little magnetic fields far below the observationally
interesting level.
The third possibility remains unresolved issue in \cite{Turner:1987bw} and several papers appeared
subsequently, investigating this case in detail.
In \cite{Garretson:1992vt}, Maxwell field coupled to pseudo Nambu-Goldstone boson $\theta$ 
by the axionic coupling $g \theta F_{\mu \nu} {\tilde F}^{\mu \nu}$,
where ${\tilde F}^{\mu \nu}$ is dual of $F_{\mu \nu}$, was studied. 
By solving the coupled equations for $A_\mu$ and $\theta$, it was shown that the resulting
magnetic field strength is tiny and is observationally irrelevant.
In \cite{Calzetta:1997ku,Kandus:1999st}, the scalar QED was studied and it was suggested
that de Sitter excitation of the charged scalar particles leads to the generation of stochastic
current on large scales. The current, after inflation, induces the magnetic field that can
be seed of the galactic dynamo.
This conclusion was revisited in \cite{Giovannini:2000dj}, taking into account
the effects of plasma conductivity. It was argued that the plasma effects suppresses the
current and was found that the resultant magnetic field is too small for seeding the
galactic magnetic field.
This issue was also addressed in \cite{Calzetta:2001cf} for the model of $N$ charged 
scalar fields coupled to the $A_\mu$ field in the large $N$ limit, 
where the magnetic field was shown to be tiny
for the chosen parameters but the generality of this result is not clear.
The acquisition of non-zero expectation value of the charged scalar field due to
the stochastic motion by the de Sitter expansion gives the gauge field an effective mass which
breaks the conformal invariance.
This effect was investigated in \cite{Davis:2000zp} by adopting the Hartree approximation
to solve the mode equation for $A_\mu$ and was found to be possible mechanism for the seed
of the galactic magnetic fields.

In \cite{Ratra:1991bn}, another possibility not covered in \cite{Turner:1987bw} was proposed,
namely, the coupling of a scalar field $\phi$(assumed to be inflaton) to the gauge kinetic term
as $e^{\alpha \phi} F_{\mu \nu}F^{\mu \nu}$, where $\alpha$ is a parameter.
Field evolution during inflation changes the magnitude of $e^{\alpha \phi}$ appearing 
in front of $F_{\mu \nu}F^{\mu \nu}$, which amounts to time variation of the gauge
coupling constant.
It was shown that it is indeed possible to generate the magnetic field strength as large as
$10^{-10}~{\rm G}$ on ${\rm Mpc}$ scales.
This type of mechanism was considered in \cite{Lemoine:1995vj} in the context of string theory
in which $\phi$ is a dilaton.
Based on that the dilaton is not suitable for realizing the potential-driven inflation 
due to its steep potential, inflation was assumed to be caused by the kinetic term of $\phi$.
Adopting the low energy effective action of heterotic string theory and putting the Kalb-Ramond field
and the axionic field to be vanishing,
solution of the mode equation for $A_\mu$ suggests that the generated magnetic field is not
strong enough to explain the observations.
The magnetic field generation in the string cosmology was also addressed in \cite{Gasperini:1995dh},
with opposite conclusion (hence positive result) to the former paper.
This is attributed to the assumption made in the latter paper of existence of the string phase.
Model of the gauge kinetic term coupled to non-inflaton $\phi$ was investigated 
in \cite{Giovannini:2001nh} for the Lagrangian density $\phi^2 F_{\mu \nu}F^{\mu \nu}$
and it was concluded that copious magnetic fields can be generated.
Dilaton's exponential coupling to $F_{\mu \nu}F^{\mu \nu}$ for the case in which inflation
is caused by the another slow-rolling scalar field was analyzed in \cite{Bamba:2003av},
taking into account the possible evolution of the dilaton until it gets stabilized.
In this case too, generation of large magnetic field was shown to be possible although
some hierarchy between the coupling to the gauge kinetic term and the steepness of the dilaton 
potential is needed to achieve the desired magnetic field strength.
Supergravity naturally allows coupling of a scalar field $\phi$ to 
$F_{\mu \nu}F^{\mu \nu}$ with arbitrary function $f(\phi)$.
Identifying such scalar field $\phi$ as inflaton, 
possibility of the magnetic field generation in the single field inflation in the 
framework of supergravity was investigated in \cite{Martin:2007ue}.
The use of the time variation of the gauge couplings constant as the generation
of the inflationary magnetic fields, which has been adopted in the papers mentioned above,
was questioned in \cite{Demozzi:2009fu}.
In that paper, it was argued that backreaction of the generated vector field on the
background severely restricts the resulting magnetic field strength and was shown that
$\sim 10^{-7}~{\rm G}$ field may be possible at the expense of the appearance of the
strong coupling epoch. 

In this paper, we do not study the modified Maxwell actions such as by giving
effective photon mass and time varying gauge coupling constant but consider the
standard quantum electrodynamics (QED) that may meet the third case
of the Turner and Widrow's classification \cite{Turner:1987bw}.
In the Standard Model of particle physics, 
electromagnetic field is coupled to charged fermions such as leptons and quarks
that have masses. 
Hence the conformal invariance in the electromagnetic field breaks down
once those charged particles enter the game.
Naively, we expect that the magnetic field coming from those charged particles 
is tiny since the violation of the conformal invariance is caused by the loop effects.
As far as we know, however, quantitative computation of the magnetic field strength
at given length scale (power spectrum) during inflation contributed from the loop
in the case of the standard QED has not been performed in the literature \footnote{
There is one exception by \cite{Dolgov:1993vg}. 
In this paper, quantum correction breaking the conformal invariance due to the 
triangle diagram with the fermion loop was investigated. 
It was found that the magnitude of this effect is negligibly small for one electron but
the effect could be significant in theories with many fermions.}.
Theoretical understanding of how much the massive charged particles (fermions) 
result in the magnetic field (on super-Hubble scales) in the inflationary Universe,
regardless of its magnitude, is a non-trivial issue, which motivates this paper.
Approximating the spacetime as pure de Sitter Universe (backreaction from the electromagnetic
field to the background is neglected), the problem is to quantize the QED on this background
and then to compute the contribution from the fermion loop to the two-point correlation 
function of the vector field.
This paper reports on the results of this calculation.
It is found that one-loop contribution coming from massive fermions compared 
to the leading tree one that respects the conformal invariance is proportional
to a product of the gauge coupling constant squared and the number of e-folds measured 
since the scale of interest crossed the Hubble radius.
Thus the one-loop contribution exhibits the secular growth, {\it i.e.} it eventually
overwhelms the tree term and the perturbative expansion breaks down. 
We then apply the dynamical renormalization technique, which effectively resums 
part of higher-loop diagrams, to show that the resummed leading loop effect enhances the 
super-Hubble scale magnetic field power spectrum.
While the tree level power spectrum scales as $\sim k^4$
($k$ is the comoving wavenumber of the power spectrum),
the loop effect changes the index of the power-law from $4$ to $4+{\cal O}(g^2) < 4$, 
where $g$ is the gauge coupling constant.
The precise form of the term ${\cal O}(g^2)$ 
dependent on the fermion mass is provided in the main part of this paper.

\section{Magnetic field on de Sitter space}
\subsection{Action of QED and its quantization}
Since spinor is not a representation of the general coordinate transformation,
contrary to the case of scalar and vector fields,
putting fermions on curved spacetime needs introduction of the tetrad field instead
of metric \cite{birrell:qgt}.
The action of QED in the general curved spacetime is given by
\begin{equation}
S= \int d^4x e \left( -\frac{1}{4} F_{\mu \nu}F^{\mu \nu}+i {\bar \Psi} \gamma^a e_a^{~\mu}D_\mu \Psi-m {\bar \Psi} \Psi  \right), \label{basic-action}
\end{equation}
where $e_a^{~\mu}$ is the tetrad field such that $g_{\mu \nu}e_a^{~\mu} e_b^{~\nu}=\eta_{ab}$ with $\eta = {\rm diag} (-1,1,1,1)$
and the covariant derivative is given by
\begin{equation}
D_\mu \Psi=(\partial_\mu+\Gamma_\mu+i g A_\mu ) \Psi,~~~~~\Gamma_\mu \equiv -\frac{1}{8} e_a^{~\nu} \nabla_\mu e_{b \nu} [\gamma^a, \gamma^b].
\end{equation}
The gamma matrices $\gamma^a$ satisfy the standard anti-commutation relations: 
$\{ \gamma^a,\gamma^b \}=-2 \eta^{ab}$.
In this paper, we work in the Dirac representation for which the $\gamma$ matrices are
written as
\begin{equation}
\gamma^0=\begin{pmatrix}
0 & {\bf I} \\
{\bf I} & 0
\end{pmatrix},~~~~~
\gamma^i=\begin{pmatrix}
0 & \sigma_i \\
-\sigma_i & 0
\end{pmatrix},
\end{equation}
where ${\bf I}$ is the $2\times 2$ unit matrix and $\sigma_i$ are the usual Pauli matrices.
The last term of the covariant derivative represents the gauge interaction with the vector field
with a coupling constant $g$.

Before going into the detailed calculations on de Sitter background,
let us briefly mention the fermion mass in the action (\ref{basic-action}).
During inflation, the Higgs field is expected to undergo the stochastic motion 
due to generation of the classical field fluctuations out of quantum fluctuations \cite{Starobinsky:1982ee, staro:1984,staro:1986}.
In the equilibrium state which is achieved after inflation lasts for ${\cal O}(100)$
$e$-folds, the Higgs field acquires the super-Hubble vacuum expectation value (VEV) 
of $\simeq \lambda^{-1/4} H$ \cite{Starobinsky:1994bd},
where $\lambda$ is the coupling constant of the Higgs self-interaction and $H$ is the
Hubble parameter during inflation.
Given that $\lambda$ does not differ significantly from unity, the Higgs field thus 
typically takes value of the order of $H$.
Since the fermions in the SM couple to the Higgs field by the Yukawa interaction,
their masses during inflation are expected to be of the order of $\simeq y H$, 
where $y$ is the Yukawa coupling.
Thus, if $y={\cal O}(1)$, then the fermion mass is the order of $H$.
Notice that we can only determine the statistically expected value of the fermion mass
due to the stochastic nature of the Higgs field and it could happen that
the actual mass is significantly deviated from the naive estimate.
In the following, we will regard $m$ as a free parameter and investigate how the
obtained magnetic field power spectrum depends on it.
Since the electroweak symmetry is broken by the VEV of the Higgs field,
we assume that the massless gauge field $A_\mu$ appearing in the 
action (\ref{basic-action}) is the usual electromagnetic field.

In the Friedmann-Lema\^itre-Robertson-Walker (FLRW) flat universe, 
the metric can be written as
\begin{equation}
ds^2=a^2 (\eta) ( -d\eta^2+\delta_{ij} dx^i dx^j ),
\end{equation}
where $\eta$ is the conformal time. 
The corresponding tetrad fields are given by
\begin{equation}
e_a^{~\mu}=\frac{1}{a} \delta^\mu_{~a}.
\end{equation}
Then the action (\ref{basic-action}) under the Coulomb gauge ($\partial_i A_i=0$) becomes
\bea
S&=&\int d\eta d^3x \left( \frac{1}{2} {\dot A_i} {\dot A_i}-\frac{1}{2} \partial_i A_j \partial_i A_j + \frac{1}{2} \partial_i A_0 \partial_i A_0 \right. \nonumber \\
&& \hspace{20mm} \left. + i a^3 {\bar \Psi} \gamma^a \partial_a \Psi+i \frac{3}{2} a^3 \frac{\dot a}{a} {\bar \Psi} \gamma^0 \Psi-a^4 m {\bar \Psi} \Psi 
-g a^3 {\bar \Psi} \gamma^a A_a \Psi \right), \label{FLRW-action}
\eea
where a dot denotes the derivative with respect to $\eta$.
The last term of \eqref{FLRW-action} represents the interaction between $A_i$ and $\Psi$ which we will treat perturbatively.

Let us first neglect the last term and quantize the free system consisting of
free vector and spinor fields.
By using the equations of motion and the remaining gauge degrees of the Coulomb gauge, 
we find that $A_0$ can be set to be zero.
With the new field $\psi$ defined by
\begin{equation}
\psi \equiv a^{3/2} \Psi,
\end{equation}
the action (\ref{FLRW-action}) then becomes
\begin{equation}
S=\int d\eta d^3x \left( \frac{1}{2} {\dot A_i} {\dot A_i}-\frac{1}{2} \partial_i A_j \partial_i A_j +
i {\bar \psi} \gamma^a \partial_a \psi- ma {\bar \psi} \psi \right).
\end{equation}
We can easily see the well-known fact that the theory becomes conformally invariant when $m=0$.
The Heisenberg equations for the field operators are reduced to
\begin{align}
{\ddot A_i}-\triangle A_i=0, \\
i \gamma^a \partial_a \psi-ma \psi=0. 
\end{align} 
We can mode-expand the field operators that satisfy the above equations as
\bea
A_i (\eta,\x) &=& \sum_s \int \f{d^3 p}{{(2\pi)}^{3/2}}\frac{1}{\sqrt{2p}} e^{i\p \cdot \x} \tilde A_i(\eta,s,\p), \\
\psi (\eta,\x) &=&  \sum_\lambda \int \frac{d^3 p}{{(2\pi)}^3} \sqrt{2}\pi^2 e^{\frac{\pi}{2} \alpha} {\left( \frac{p}{Ha} \right)}^{1/2} e^{i\p \cdot \x} \tilde \psi (\eta,\lambda,\p), 
\eea
where 
\bea
\tilde A_i(\eta,s,\p) &=& \e_i(\p,s) \mk{e^{-ip\eta}c(\p,s)+e^{ip\eta}c^\dagger(-\p,s)} ,\\
\tilde \psi (\eta,\lambda,\p) &=& u(\p,\eta;\lambda)a(\p,\lambda) + v(-\p,\eta;\lambda) b^\dagger(-\p,\lambda),
\eea
and the sum over $s$ and $\lambda$ for each field is taken to account for the two different spin states 
and $\alpha \equiv m/H$.
The polarization vector $\e_i(\p,s)$ satisfies following equations:
\bea
p_i\e_i(\p,s)&=&0, \\
\e_i(\p,s_1)\e_i(\p,s_2)&=&\d_{s_1 s_2}, \\
\sum_{s} \e_i(\p,s)\e_j(\p,s)&=&\d_{ij}-\f{p_ip_j}{p^2}.
\eea
The four-component spinors $u$ and $v$ are defined by \cite{Cotaescu:2001cv},
\begin{align}
u({\vec p},\eta;\lambda) &=\begin{pmatrix}
H_{1/2-i\alpha}^{(1)} \left( \frac{p}{Ha} \right) \xi_\lambda ({\vec p}) \\
\lambda e^{-\pi \alpha}H_{1/2+i\alpha}^{(1)} \left( \frac{p}{Ha} \right) \xi_\lambda ({\vec p})
\end{pmatrix},\\
v({\vec p},\eta;\lambda) &=i \gamma^2 u^* ({\vec p},\eta;\lambda)= \begin{pmatrix}
\lambda e^{-\pi \alpha} H_{1/2-i\alpha}^{(2)} \left( \frac{p}{Ha} \right) \theta_\lambda ({\vec p}) \\
-H_{1/2+i\alpha}^{(2)} \left( \frac{p}{Ha} \right) \theta_\lambda ({\vec p}), 
\end{pmatrix}.
\end{align}
Here $\xi_\lambda ({\vec p})$ is defined by an equation:
\begin{equation}
\frac{{\vec \sigma} \cdot {\vec p}}{p}\xi_\lambda ({\vec p})=\lambda \xi_\lambda ({\vec p}),
\end{equation}
and is normalized to unity.
The eigenvalue $\lambda$ is either $1$ or $-1$. $\theta_\lambda ({\vec p})$ is defined by $\theta_\lambda ({\vec p}) \equiv i \sigma_2\xi^*_\lambda ({\vec p})$ and satisfies an equation:
\begin{equation}
\frac{{\vec \sigma} \cdot {\vec p}}{p}\theta_\lambda ({\vec p})=-\lambda \theta_\lambda ({\vec p}).
\end{equation}
The norm of $\theta_\lambda ({\vec p})$ is also unity.

We impose the canonical quantization conditions:
\begin{align}
[A_i (\eta,{\vec x}),\Pi_j (\eta,{\vec y})]&=i \int \frac{d^3 p}{{(2\pi)}^3}
\left( \delta_{ij}-\frac{p_i p_j}{p^2}\right) e^{i {\vec p} \cdot ({\vec x}-{\vec y})}, \\
\{ \psi_a (\eta,{\vec x}), \pi_b (\eta,{\vec y}) \}&= i \delta_{ab} \delta ({\vec x}-{\vec y}),
\end{align}
where $\Pi_i = \partial_\eta A_i$ and $\pi_a =i \psi_a^*$ are canonical conjugate momenta
of the vector and the spinor fields, respectively. 
All the other (anti-)commutation relations are zero.
We can check that these conditions are derived by the following (anti-)commutation relations:
\begin{align}
&\{ a ({\vec p_1},\lambda_1),a^\dagger ({\vec p_2},\lambda_2) \}= \{ b({\vec p_1},\lambda_1),b^\dagger ({\vec p_2},\lambda_2) \}=\delta_{\lambda_1 \lambda_2} \delta ({\vec p_1}-{\vec p_2}),\\
&[c({\vec p_1},s_1),c^\dagger ({\vec p_2},s_2)]=\delta_{s_1 s_2} \delta ({\vec p_1}-{\vec p_2}),
\end{align}
and all the other (anti-)commutation relations are zero.

\subsection{Magnetic field power spectrum}
Before going into the main calculations, let us first define the magnetic
field (magnetic flux) power spectrum.
Physical magnetic flux density is defined by
\bea
B_i(\eta,\x) &=& a^{-2}\e_{ijk} \partial_jA_k(\eta,\x).
\eea
The corresponding Fourier component ${\tilde B}_i$ is given by
\bea
{\tilde B}_i (\eta,{\vec k})=\int d^3x~e^{-i{\vec k}\cdot {\vec x}}B_i(\eta,\x).
\eea
Then the power spectrum of the magnetic field is defined by
\begin{equation}
\langle {\tilde B}_i (\eta,{\vec k_1}) {\tilde B}_i (\eta,{\vec k_2}) \rangle
={(2\pi)}^3 \delta ({\vec k_1}+{\vec k_2}) P_B(k_1),
\end{equation}
where $\langle \cdots \rangle$ is the ensemble average which, as usual,
is identified with the vacuum expectation value.
We also introduce another spectrum ${\cal P}_B(k)$ defined by
\begin{equation}
{\cal P}_B(k)=\frac{k^3}{2\pi^2} P_B(k),
\end{equation}
which represents power of $B$ per logarithmic interval of $k$.
Using these definitions, ${\tilde B}_i$ is found to be related to ${\tilde A}_i$ as
\begin{equation}
{\tilde B}_i (\eta,{\vec k})=\frac{i {(2\pi)}^{3/2}}{a^2 \sqrt{2k}}
\epsilon_{ijk}k_j \sum_s {\tilde A_k} (\eta,s,{\vec k}).
\end{equation}

Having quantized the free part of the Hamiltonian, it is straightforward to 
calculate the expectation value of any operator consisting of field operators.
In the in-in formalism,
the expectation value of an operator $Q(t)$ in the interaction picture is given by
({\it e.g}., \cite{Weinberg:2005vy})
\begin{equation}
\langle Q(t) \rangle = \bigg\langle \bigg[ {\bar T} \exp \left( i \int_{t_0}^t dt' ~H_I (t') \right) \bigg] Q_I (t) \bigg[ T\exp \left( -i \int_{t_0}^t dt' ~H_I (t') \right) \bigg] \bigg\rangle,
\end{equation}
where $T$ and ${\bar T}$ are time-ordering and anti-time-ordering operators.
$H_I$ is the interaction Hamiltonian in which all the fields are given in the interaction picture.
There is another equivalent expression of the above expectation value:
\begin{align}
\langle Q(t) \rangle =& \sum_{N=0}^\infty i^N \int_{t_0}^t dt_N~\int_{t_0}^{t_N} dt_{N-1}~\cdots \int_{t_0}^{t_2}dt_1 \notag \\
& \bigg\langle \bigg[ H_I (t_1),~\bigg[ H_I(t_2),\cdots \bigg[ H_I (t_N),Q_I (t) \bigg] \cdots \bigg] \bigg] \bigg\rangle. 
\end{align}
In our case, the interaction Hamiltonian is given by
\footnote{To be precise, $H_I(t)$ includes another term. 
The equation of motion for $A_0$ derived from Eq.~\eqref{FLRW-action} is 
$\del_i\del_i A_0 + g a^3 \bar \Psi \gamma^0 \Psi = 0$. 
Therefore, by substituting the solution of $A_0$ back into Eq.~\eqref{FLRW-action}, 
we have another interaction term, which is of the order of $g^2\psi^4$. 
This contributes $H_I(t)$. 
However, its contribution to the two-point correlation function of the magnetic field
appears at the order of $g^4$ while contribution coming from Eq.~(\ref{HI}) is
the order of $g^2$. Thus, we do not include it in \eqref{HI}. }
\begin{equation}
H_I(t) = g \int d^3x~{\bar \psi} \gamma^a A_a \psi. \label{HI}
\end{equation}
In terms of mode expanded fields,
\be H_I(\eta)=\f{2\pi^4e^{\pi\alpha}g}{Ha(\eta)} 
\sum_{\lambda,\mu,w} \fm{k}\fm{\ell}\int\frac{d^3h}{{(2\pi)}^{3/2}} \sqrt{\f{k\ell}{2h}} 
(2\pi)^3\d(-\k+\l+\h) \tilde H_I(\eta,\lambda,\mu,w,\k,\l,\h), \ee
with
\be
\tilde H_I(\eta,\lambda,\mu,w,\k,\l,\h) = \tilde\psi^\dagger(\eta,\lambda,\k)\g^0\g^i\tilde A_i(\eta,w,\h)\tilde\psi(\eta,\mu,\l).
\ee
The two-point function of the magnetic field is then given by 
\bea 
\langle {\tilde B}_i (\eta,{\vec k_1}) {\tilde B}_i (\eta,{\vec k_2}) \rangle
&=&\sum_{N=0}^{\infty}i^N \int_{-\infty}^\eta d\eta_N \int_{-\infty}^{\eta_N} d\eta_{N-1} \cdots \int_{-\infty}^{\eta_2} d\eta_1 \notag \\
&& \vac{\kk{H_I(\eta_1),\kk{H_I(\eta_2),\cdots,\kk{H_I(\eta_N),
{\tilde B}_i (\eta,{\vec k_1}) {\tilde B}_i (\eta,{\vec k_2})}\cdots } } }, \notag \\
&\equiv& {(2\pi)}^3 \delta ({\vec k_1}+{\vec k_2}) \sum_{N=0}^{\infty} P_B^{(N)}(\eta,k). \label{2-pt}
\eea
Note that $P_B^{(N)}$ is of the order of $g^N$ because $H_I$ is proportional to $g$.

For the purpose of the completeness, let us first give the zeroth order term.
After some calculations, we find
\begin{equation}
P_B^{(0)}(\eta,k)=\frac{k}{a^4(\eta)},~~~~~{\cal P}_B^{(0)}(\eta,k)=\frac{k^4}{2\pi^2 a^4(\eta)},
\end{equation}
which recovers the well-known result $P_B \propto a^{-4}$.

\begin{figure}[t]
\centering
\includegraphics[width=120mm]{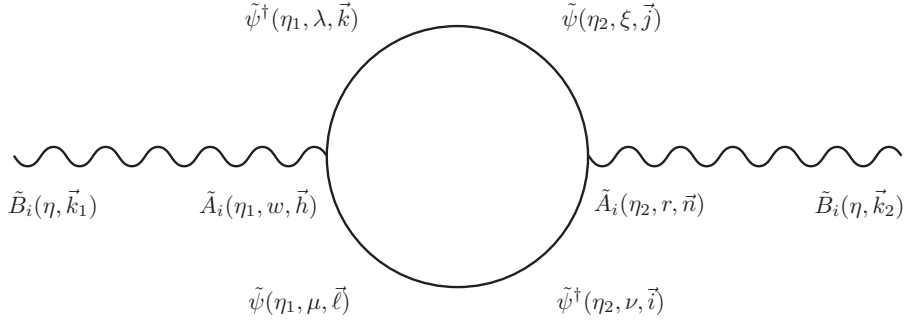}
\caption{
The one-loop diagram corresponding to Eq.~\eqref{Psec}.
}
\label{fig:diag}
\end{figure}

The first order term vanishes because it includes odd number of creation/annihilation operators.
The first nonzero contribution comes from the second order term. 
Substituting mode expansion of interaction Hamiltonian, we find that the term 
second order in $g$ in Eq.~(\ref{2-pt}) can be written as
\bea
&&- \int_{-\infty}^\eta d\eta_2 \int_{-\infty}^{\eta_2} d\eta_1~
\f{2\pi^4e^{\pi\alpha}g}{Ha(\eta_1)} \sum_{\lambda,\mu,w} 
\int \frac{d^3k}{{(2\pi)}^3} \frac{d^3\ell}{{(2\pi)}^3}\frac{d^3h}{{(2\pi)}^{3/2}} 
\sqrt{\f{k\ell}{2h}} (2\pi)^3\d(-\k+\l+\h) \notag \\
&& \times \f{2\pi^4e^{\pi\alpha}g}{Ha(\eta_2)} \sum_{\nu,\xi,r} 
\int \frac{d^3i}{{(2\pi)}^3} \frac{d^3j}{{(2\pi)}^3} \frac{d^3n}{{(2\pi)}^{3/2}}
\sqrt{\f{ij}{2n}} (2\pi)^3\d(-\ii+\j+\n) \notag \\
&& \times \vac{\kk{\tilde H_I(\eta_1,\lambda,\mu,w,\k,\l,\h),\kk{\tilde H_I(\eta_2,\nu,\xi,r,\ii,\j,\n),\tilde B_i(\eta,{\vec k_1})\tilde B_i(\eta,{\vec k_1})}}}.
\label{Psec}
\eea
Manipulations of the commutation relations in this expression yield 
terms which are proportional to $\d({\vec k_1})\d({\vec k_2})$ and 
$\d({\vec k_1}+{\vec k_2})$, respectively. 
Since we are interested in the inhomogeneous magnetic fields (finite coherent length), 
we consider only the latter contribution.
Alternatively, we can apply the diagrammatic technique to evaluate Eq.~(\ref{Psec}).
In this approach, we construct a diagram corresponding to Eq.~(\ref{Psec}) (represented in Fig.~\ref{fig:diag}), 
assign suitable quantity to each vertex, external and internal line and
take a product of them (for a concrete prescription, see \cite{Weinberg:2005vy}).
In either way, after integration and taking sum (straightforward but cumbersome), we obtain 
\bea
{\cal P}_B^{(2)}(\eta,k)
&=& \f{e^{2\pi \a}}{a^4}\f{g^2}{64\pi^3} \int_{-\infty}^\eta d\eta_2 
\int_{-\infty}^{\eta_2} d\eta_1~k \eta_1\eta_2 \int d^3p~ \notag \\
&& {\rm Re} \kk{ \mk{e^{ik(-\eta_1+\eta_2)}-e^{ik(2\eta-\eta_1-\eta_2)}} 
\ck{|\p-\k|p k^2 {\cal I}_1 + \mk{(\p-\k)\cdot \k} (\p\cdot \k) {\cal I}_2 } },
\label{rho2p7}
\eea
where
\bea
{\cal I}_1&=&e^{-4\pi\a} \Hfp(-p\eta_1) \Hfp(-|\p-\k|\eta_1) \Hsm(-p\eta_2) \Hsm(-|\p-\k|\eta_2) \notag \\
&&+ \Hfm(-p\eta_1) \Hfm(-|\p-\k|\eta_1) \Hsp(-p\eta_2) \Hsp(-|\p-\k|\eta_2), \\
{\cal I}_2&=&e^{-2\pi\a}\biggl[ \Hfp(-p\eta_1) \Hfp(-|\p-\k|\eta_1) \Hsp(-p\eta_2) \Hsp(-|\p-\k|\eta_2)  \biggr. \notag \\
&&+\biggl. \Hfm(-p\eta_1) \Hfm(-|\p-\k|\eta_1) \Hsm(-p\eta_2) \Hsm(-|\p-\k|\eta_2) \biggr].
\eea

\begin{figure}[t]
\centering
\includegraphics[width=100mm]{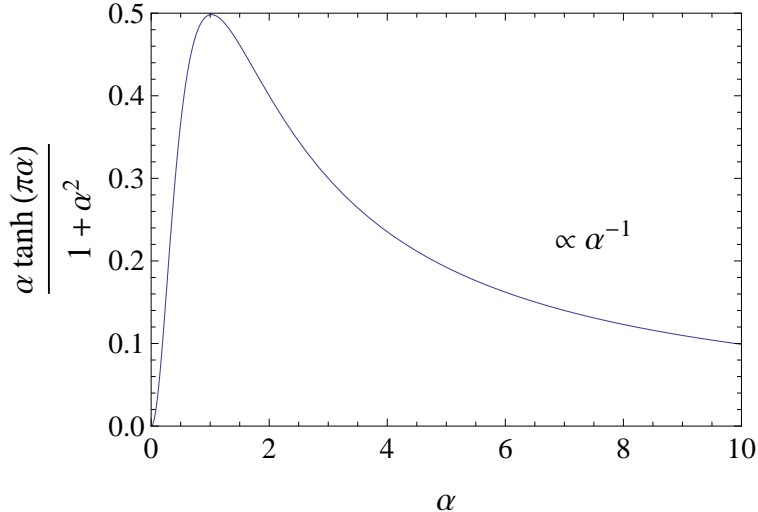}
\caption{
The dependence on $\alpha\equiv m/H$ of the one-loop contribution from the super-Hubble modes to the power spectrum ${\cal P}_B^{(2)}(\eta,k)$ given in Eq.~(\ref{1-loop-result}).
The maximum value is $\simeq 0.498$ at $\alpha \simeq 1.02$.
For $\alpha\to \infty$, ${\cal P}_B^{(2)}(\eta,k)$ scales as $\alpha^{-1}$.}
\label{fig:alpha}
\end{figure}

So far, we have not used any approximation to derive Eq.~\eqref{rho2p7}. 
As is clear from the expressions for ${\cal I}_1$ and ${\cal I}_2$,
their complex structures defeat exact integration of Eq.~\eqref{rho2p7}. 
Nevertheless, the infrared (IR) secular growth, which is the most important ingredient in our analysis, 
can be analytically evaluated by making the following approximations.
Firstly, we remove angular dependence from $|\k-\p|$ and replace it with $|k-p|$. 
We then take angular integral both for $k$ and $p$. 
Secondly, we neglect the contribution from the sub-Hubble modes,
{\it i.e.}, the domain where arguments of the Hankel functions become larger than unity \footnote{
It can be checked that Eq.~\eqref{rho2p7} is ultraviolet (UV) divergent and its 
contribution takes a form of ${\cal P}_B \supset C/a^4$, where $C$ is a divergent constant.
As usual, such a divergence can be eliminated by renormalization of the gauge field $A_\mu$.
In other words, the divergent term can be canceled with the contribution from
a counter term of a form: $-(Z-1) \frac{1}{4} F_{\mu \nu}F^{\mu \nu}$. 
After renormalization, finite contribution to ${\cal P}_B$ of a form
$C'/a^4$ remains, where $C'$ is a finite constant.
On the other hand, the IR contribution to ${\cal P}_B$ takes a form of
Eq.~(\ref{1-loop-result}) which is enhanced by $\ln a$ for $a \gg 1$ 
({\it i.~e.}, sufficiently late time) compared to the UV contribution.
}. 
In other words, we take integrations only when the following inequalities are satisfied:
\be p-\f{1}{|\eta_2|}\leq k \leq \f{1}{|\eta_2|},\quad p-\f{1}{|\eta_1|}\leq k \leq \f{1}{|\eta_1|}. \label{kineq} \ee
In order for positive $k$ to exist that satisfies the above inequalities \eqref{kineq}, 
the interval for each inequality should be positive. 
This condition determines the integration range for $\eta_1$ and $\eta_2$:
\be \int^\eta_{-\infty}d\eta_2 \int^{\eta_2}_{-\infty}d\eta_1 \to \int^\eta_{-\f{2}{p}}d\eta_2 \int^{\eta_2}_{-\f{2}{p}}d\eta_1\ee
From $|\eta_1|>|\eta_2|>|\eta|$, the range for $k$-integration becomes
\be \int^\infty_0 dk \to \int^{\f{1}{|\eta_1|}}_{{\rm Max} \ck{0,~ p-\f{1}{|\eta_1|}} } dk. \ee
Furthermore, we take super-Hubble limit $-p\eta \ll 1$, 
and use an asymptotic form of Hankel function. For $|z|\ll 1$,
\be H^{(1,2)}_\nu(z) \sim \mp \f{i\Gamma(\nu)}{\pi}\mk{\f{z}{2}}^{-\nu}. \ee
Finally, we only keep the lowest order of $-p\eta_1$ and $-p\eta_2$ and 
evaluate the contribution from the super-Hubble modes. 
The result is given by
\be
{\cal P}_B^{(2)}(\eta,k) \simeq -\frac{g^2}{3\pi^4}\f{\a \tanh (\pi\a)}{1+\a^2} \frac{k^4}{a^4} 
\ln \left(\frac{k}{aH} \right). \label{1-loop-result}
\ee
The $\alpha$-dependence is shown in Fig.~\ref{fig:alpha}.
As is clear from its derivation, this result is applicable to the super-Hubble 
modes $k \ll aH$.
For the super-Hubble modes, the one-loop effect gives positive contribution 
to the magnetic field power spectrum, thus enhancing the power in the infrared regime.
We also find Eq.~(\ref{1-loop-result}) is an even function of $\alpha =m/H$.
This is consistent with that only the absolute value of $m$ has
a physical meaning for fermions.
We also find that Eq.~(\ref{1-loop-result}) vanishes in the massless limit 
$\a\to 0$.
This lack of infrared enhancement in the massless case may be interpreted as the 
conformal invariance of the total action.
For fixed $g,~H$ and $k/a$, the one-loop correction takes a maximum at $\alpha \simeq 1.02$ and monotonically decreases for larger $\alpha$.
In particular, the massive limit $\alpha \to \infty$ gives vanishing ${\cal P}_B^{(2)}$.
This is consistent with the standard picture that excitation of particles 
heavier than $H$ in de Sitter space is suppressed on super-Hubble scales.
To summarize, the one-loop contribution becomes maximal when the fermion mass
is close to the Hubble parameter of inflation.

Combining the zeroth order contribution, the magnetic field spectrum 
(on super-Hubble scales) up to one-loop order is given by
\begin{equation}
{\cal P}_B(\eta,k) \simeq \frac{k^4}{2\pi^2 a^4} \bigg[ 
1-\frac{2g^2}{3\pi^2} \f{\a \tanh (\pi\a)}{1+\a^2}\ln \left(\frac{k}{aH} \right) \bigg]. \label{fin-1loop}
\end{equation}
Let us estimate the relative contribution of the one-loop term evaluated at our
current cosmological scale $\sim 1~{\rm Gpc}$ which corresponds to  
$\ln \left(\frac{k}{aH} \right) \sim -65$
(precise value depends on the inflation models).
Choosing $\alpha=1.02$ in order to estimate the possible maximum value of ${\cal P}_B$,
we find that the second term in the square bracket of Eq.~(\ref{fin-1loop}) is
\begin{equation}
-\frac{2g^2}{3\pi^2} \f{\a \tanh (\pi\a)}{1+\a^2}\ln \left(\frac{k}{aH} \right) 
\simeq 0.35 ~{\left( \frac{g}{0.4} \right)}^2,
\end{equation}
where we have used the running gauge coupling constant 
at the energy scale $10^{15}~{\rm GeV}$.
This shows that the one-loop effect enhances the magnetic field power by at most $35~\%$.
Since the zeroth order term is known to give extremely tiny magnetic field 
from the observational point of view \cite{Turner:1987bw}, 
mere $35~\%$ enhancement does not yield the observationally relevant magnetic field strength.
Therefore, we conclude that the loop corrections are not capable of producing
the seed magnetic field at least in the simple model we consider in this paper.

On the other hand, if there are multiple fermionic fields, 
each field adds the similar contribution to ${\cal P}_B$.
As a result, the net contribution becomes simply a sum over each contribution at one-loop order. 
In other words, adding the fermionic field enhances the loop effect.
Additionally, increase of the gauge coupling constant also amplifies the loop effect.
Currently, we do not know how many light fermionic fields coupled to the electromagnetic field exist 
and how strong the gauge coupling constant is at the energy scale of inflation which is also unknown.
Thus the possibility remains that loop effect generates large magnetic field strength during inflation.

\subsection{Secular growth and resummation}
Equation~(\ref{fin-1loop}) shows that the one-loop contribution exhibits the infrared secular growth,
{\it i.e.,} the second term in the square bracket of Eq.~(\ref{fin-1loop}) grows
in proportion to the number of $e$-folds measured since the mode of interest crossed
the Hubble horizon.
This implies that if inflation lasts sufficiently long time, the one-loop term eventually
dominates over the tree term and the perturbative calculation implemented in this paper
becomes no longer valid.
In such a situation, despite the coupling constant is small,
the higher loop effects need to be taken into account to correctly estimate the power spectrum.
The appearance of the secular growth for the infrared correlation functions on de Sitter space
is generic in many models.
For instance, for the minimally coupled massless scalar field $\phi$ having quartic self-interaction
${\cal L}_{\rm int}=-\frac{\lambda}{4}\phi^4$,
it is known that equal-time two-point function of $\phi$,
when it is computed by means of the perturbative expansion in terms of the coupling constant $\lambda$,
exhibits the infrared secular growth proportional to the number of $e$-folds, 
which is exactly the same behavior as Eq.~(\ref{fin-1loop})
(for instance, see \cite{Sasaki:1992ux,Suzuki:1992gi,Seery:2010kh,Tanaka:2013caa}).

The breakdown of the perturbative expansion even for the small coupling constant reminds us
of the similar situation met in the computation of the scattering amplitude in the framework
of the standard quantum field theory (in Minkowski space). 
At high energies, the one-loop scattering amplitude is generically accompanied with $\ln q$ 
($q$ is the energy scale of interest) in addition to the coupling constant which is assumed to be 
sufficiently small.
Thus, if we go beyond a certain energy scale, 
$\ln q$ becomes so large that the one-loop correction
is no longer suppressed compared to the tree term. 
The method to handle this problem is the renormalization group (RG) technique
which effectively absorbs the higher loop contributions partially into the coupling constant.
In terms of the newly defined coupling constant which now depends on the energy scale,
the scattering amplitude is free from the $\ln q$ factor and the perturbative expansion
works as long as the coupling constant is small.
Now, if we identify $\ln q$ with $\ln (k/aH)$ in Eq.~(\ref{fin-1loop}), the issue of the
secular growth is apparently similar to the $\ln q$ enhancement in the scattering amplitude.

In \cite{Burgess:2009bs}, the use of the dynamical RG method was proposed to tame the 
infrared secular growth in de Sitter space.
Although the basic idea of the dynamical RG is the same as the usual RG,
one difference is that while the higher loop effects are pushed into the coupling constant 
in the case of the standard RG, it is the power spectrum that plays this role in the case
of the dynamical RG.
It was also demonstrated in \cite{Burgess:2009bs} that, 
for the massless scalar field with quartic self-interaction,
the two-point function after the dynamical RG is applied becomes less singular 
at infrared than the one-loop effect only and recovers the one obtained by
the resummation of the chain diagrams \cite{Petri:2008ig}.

Now, let us apply the dynamical RG method to Eq.~(\ref{fin-1loop}) to make it regular
even at far infrared regime.
To this end, let us first interpret Eq.~(\ref{fin-1loop}) that it connects the magnetic field 
power spectrum at horizon crossing time with the one at sufficiently later time. 
Then, following \cite{Burgess:2009bs}, this equation may be interpreted that the (leading) loop effect
always enhances the power spectrum by multiplying the power spectrum by the amount 
\begin{equation}
f=1+\frac{2g^2}{3\pi^2} \frac{\alpha \tanh (\pi \alpha)}{1+\alpha^2} N,
\end{equation}
after the universe expands by $N$ $e$-folds.
But, this procedure is reliable only when $N$ is small enough so that the loop correction
in the factor $f$ in the above equation is smaller than unity.
Then, in order to obtain the power spectrum at very late time ({\it i.e.}, large $N$),
we need to divide the time interval between the final time and the horizon crossing time
into short pieces and we apply the above procedure of multiplying by the factor $f$
at each interval for which such a procedure is a good approximation.
Denoting the time interval of one piece by $\Delta \eta$ (in conformal time),
the power spectrum at $\eta+\Delta \eta$ may be written as
\begin{equation}
{\cal P}_B (k,\eta+\Delta \eta)=\frac{a^4(\eta)}{a^4(\eta+\Delta \eta)} {\cal P}_B(k,\eta)
\bigg[ 1+\frac{2g^2}{3\pi^2} \frac{\alpha \tanh (\pi \alpha)}{1+\alpha^2} \ln \left( \frac{a(\eta+\Delta \eta)}{a(\eta)}\right) \bigg].
\end{equation}
Now, making $\Delta \eta$ infinitesimal, this equation becomes a differential equation whose
solution can be obtained immediately;
\begin{equation}
{\cal P}_B (k,\eta)=\frac{H^4}{2\pi^2} {\left( \frac{k}{aH} \right)}^{4-\nu}, \quad \quad
\nu=\frac{2g^2}{3\pi^2} \frac{\alpha \tanh (\pi \alpha)}{1+\alpha^2}. \label{exp-nu}
\end{equation}
It is clear that the secular growth disappears in the resummed power spectrum 
and the loop effect acts in decreasing the power of $k/(aH)$ from 4 (tree level)
by $\nu >0$.
Thus, the loop effect makes the power spectrum redder compared to the tree level one.
The expression of $\nu$ given above represents the leading loop effect.
The higher order effects also contribute to $\nu$, which is ${\cal O}(g^4)$.
Explicit computation of such term is beyond the scope of this paper.

Let us evaluate the magnitude of $\nu$.
Assuming $N_\psi$ fermion fields having masses equal to $H$ ($\alpha=1$) as an optimistic estimation,
we find
\begin{equation}
\nu = 0.054~{\left( \frac{g}{0.4} \right)}^2 \left( \frac{N_\psi}{10} \right). \label{value-nu}
\end{equation}
In order for the loop effect to be responsible for the seed of the observed magnetic fields,
$\nu \gtrsim 1$ is at least required.
Therefore, as discussed in the previous subsection,
either large gauge coupling constant or large number of fermion fields is necessary for that purpose.

\section{Summary}
Observed magnetic fields in the universe might have been generated during the primordial
inflationary epoch.
This possibility has been investigated intensively in the literature from various point of view.
That the Maxwell action without interactions is conformally invariant implies that the magnetic 
power spectrum generated during inflation scales as $k^4/a^4$ ($k$ and $a$ is wavenumber and the scale factor, respectively),
resulting in extremely tiny magnetic field strength that has no observational relevance. 
This scaling can be modified in the presence of interactions.

In this paper, we computed the effect of the vacuum polarization to the infrared magnetic field 
power spectrum on de Sitter background in the framework of the standard QED
in which the conformal invariance is broken by the massive fermion coupled to the electromagnetic field.
After having reviewed the quantization procedure of both fermion and vector fields on
de Sitter space,
we explicitly implemented the calculation of the power spectrum up to the lowest
non-trivial order (one-loop) of the gauge coupling constant by means of the in-in formalism.
We found that the one-loop term exhibits the secular growth at infrared regime, which grows in proportion
to the number of $e$-folds measured since the mode of interest crossed the Hubble horizon, 
compared to the leading tree term.
In order to make the loop correction regular,
we applied the dynamical renormalization group method which amounts to the partial resummation
of the higher order terms. 
The resummed power spectrum was found to be free from the secular growth and the loop effect
only changes the power as ${(k/a)}^{4-\nu}$, where $\nu$ depends on the gauge coupling constant
as well as the fermion mass and its explicit expression is given by Eq.~(\ref{exp-nu}).
The $\nu$ is non-negative and thus the loop effect always enhances the infrared magnetic field strength.
For fixed gauge coupling constant, $\nu$ becomes maximum when the fermion mass $m$ becomes $1.02~H$
($H$ is the Hubble parameter).
We also confirmed that $\nu$ vanishes in the massless limit of the fermion in which case
the full action including the interaction term respects the conformal invariance and
in the massive limit.
Choosing the fermion mass to be equal to the Hubble parameter, we found $\nu \simeq 5 \times 10^{-3}$
for the gauge coupling constant $g=0.4$.
Since $\nu \gtrsim 1$ is at least needed to explain the observed magnetic fields,
this estimate suggests that the loop effect from a single fermion with the gauge coupling constant of 
$g < 1$ is not enough to be responsible for the required seed magnetic fields.
On the other hand, if the gauge coupling constant at the inflation energy scale is much larger than 
that at the electroweak scale
or if there exist so many fermion fields coupled to the electromagnetic field,
$\nu$ can be much larger the above estimate and the loop effect may be able to explain the
observed magnetic field.
Although there is no strong support for such possibilities from particle physics, 
they are not completely precluded.\\

\noindent {\bf Acknowledgments:} 
This work is supported by Grant-in-Aid for Scientific Research on Innovative Areas
No.~25103505 (TS) from The Ministry of Education, Culture, Sports, Science and Technology (MEXT) and JSPS Postdoctoral Fellowships for Research Abroad (HM).

\bibliographystyle{JHEP}
\bibliography{draft}

\end{document}